\documentstyle[aps,epsfig,psfig]{revtex}
\begin {document}
\title {Tensionless structure of glassy phase} 
\author{A.~Lipowski$^{1)}$ and D.~Johnston$^{2)}$}
\address{
$^{1)}$ Department of Physics, A.~Mickiewicz University,
61-614 Pozna\'{n}, Poland\\
$^{2)}$ Department of Mathematics, Heriot-Watt University
EH14 4AS Edinburgh, United Kingdom\\}
\maketitle
\begin {abstract}
We study a class of homogeneous finite-dimensional Ising models which 
were recently
shown to exhibit glassy properties.
Monte Carlo simulations of 
a particular three-dimensional model in this class
show that the
glassy phase obtained under slow cooling is dominated by large scale
excitations whose energy $E_l$ scales with their size $l$ as $E_l\sim
l^{\Theta}$ with $\Theta\sim 1.33(5)$. 
Simulations suggest that in another model of this class, namely the
four-spin model, energy is concentrated mainly
in linear defects making also in this case domain walls tensionless.
Two-dimensinal variants of these models are trivial and
energy of excitations scales with the exponent $\Theta=1.05(5)$.
\end{abstract} \pacs{}
Recently the problem of the structure of the glassy phase in spin glasses
has attracted considerable attention.
The main thrust of research has been to establish whether the low-temperature
phase is described by the so-called "droplet model"~\cite{DROPLET} or by the
replica symmetry breaking (RSB) theory of Parisi~\cite{RSB}.
An important difference between these theories concerns the energy $E_l$
of large-scale excitations which should scale with their size $l$ as 
$E_l\sim l^{\Theta}$ where $\Theta>0$ for the droplet model but with
$\Theta=0$ for the RSB theory.
The most interesting situation arises in the three-dimensional case,
where it seems that a combination of these two approaches is needed to
describe the glassy phase correctly~\cite{MARTIN}.

Of course, the problem of the structure of the glassy phase is not
restricted to spin glasses.
Disordered, out of equilibrium slow dynamics structures appear in
superconducting compounds, polymers and granular matter.
However, these systems are very complex and modelling them seems
to be more difficult than understanding spin glasses.
Conventional glasses are also very complex~\cite{GOTZE}.
Nevertheless, recently relatively simple models have been
proposed which exhibit an encouraging
number of glassy properties~\cite{SHORE,LIPDES,NEWMAN,FRANZ,BERNASCONI}.
All these models are spin models which do not contain quenched
disorder as in the case of spin glasses and glassiness is dynamically
generated.
An absence of quenched disorder has important implications.
Firstly, their ground state
and sometimes even the structure of excitations are known exactly.
Let us emphasize that for the three-dimensional spin glasses the problem of
finding the ground state is extremely difficult (NP-complete) and is one of
the main difficulty in numerical approaches to spin glasses.
Secondly, for models without quenched disorder there is no need to average over
different realizations of this disorder, which is yet another computational
advantage of such models.

The objective of the present paper is to examine the nature of glassy
phase in certain nondisordered Ising models.
Simple 
heuristic
arguments show that these models might have large scale excitations of
energy which scale with their size as $l^{d-2}$, i.e., slower than their
surface ($\sim l^{d-1}$). 
The question is whether such states appear in, or maybe even
dominate, the glassy phase.
To examine this problem we have
performed Monte Carlo simulations of the models and our
results suggest that the glassy phase is dominated by excitations whose energy
increases faster than $l^{d-2}$ but slower than $l^{d-1}$.
It is also likely that the exponent $\Theta$ which describes the size
dependence of energy of excitations takes some nontrivial values for these
models.
For the models of conventional glasses 
considered here the problem
of energetics of large scale excitations is computationally much more
tractable than for spin glass models.
Hopefully, results obtained for these models will provide some insight into
other glassy systems too.
In addition, our results can be used to verify some earlier claims
concerning
the nature of the glassy transition in some of these
models~\cite{LIPDESESPRIU}.

The class of models which we examine in this paper is defined by the following
Hamiltonian 
\begin{equation}
H=-2k\sum_{<i,j>} S_iS_j +\frac{k}{2}\sum_{<<i,j>>} S_iS_j
+\frac{(1-k)}{2}\sum_{[i,j,k,l]} S_iS_jS_kS_l, \label{e1}
\end{equation}
where summations in~(\ref{e1}) are over nearest neighbours, next-nearest
neighbours, and elementary plaquettes respectively.
Model~(\ref{e1}) has the interesting property that the energy of
cerain excitations of size $l$ is proportional to $l^{d-2}$, whereas
typically the energy of an excitation for 
a standard nearest neighbour Ising model is proportional to its
surface ($\sim l^{d-1}$).)
This property has been used to construct a class of
random surface
theories based on model~(\ref{e1})~\cite{SAVVIDY}.  

Recently, it was shown that for $d=3$ model (\ref{e1}) has slow dynamics
at low temperature~\cite{LIPDESESPRIU}.
When the high temperature sample is quenched to low temperature the
excess energy $\delta E=E(t)-E_{eq}$, where $E_{eq}$ is the equilibrium energy,
decays with time $t$ much more slowly than $t^{-1/2}$, which is a typical
decay rate
for Ising
models with nonconservative dynamics~\cite{BRAY}.
It turns out that 
the 
$k=0$ (pure four-spin interaction) case is 
of particular interest.
This is because in this case the model has also some other
properties typical of conventional glasses such as strong
metastability~\cite{LIPDES} and small cooling-rate effects~\cite{LIPDES2}.  
Moreover, certain time dependent
correlation functions, such as those describing aging, also behave
similarly to real glassy systems~\cite{SWIFT}.
Although a slow decay of $\delta E$ is an indication of slow dynamics, 
it would
be desirable to relate this decay with the increase of a characteristic length
scale $l$.
(As we will see such a relation will give some information about the
energetis of excitations of the glassy phase.)

For ordinary Ising models simple arguments, based on the fact that the energy
of excitation of the size $l$ scale as its surface ($E_l \sim l^{d-1}$),
can be
used to obtain the relation
\begin{equation}
\delta E\sim \frac{1}{l}.
\label{e2}
\end{equation}
However, since for model~(\ref{e1}) energy of excitations might increase 
more slowly
than their surface area,  the relation~(\ref{e2}) is no longer obvious.
Assuming that in the glassy phase the dominant
excitations are indeed these low-energy excitations (with $E_l \sim
l^{d-2}$), the following relation
should hold~\cite{LIPDES2}
\begin{equation}
\delta E\sim \frac{1}{l^{2}}.
\label{e3}
\end{equation}
Let us notice that the assumption that the glassy phase is composed of
low-energy excitations implies that at the glassy transition
domain walls lose their surface tension.
Such an identification might be of more general validity and could be used as a
new criterion to locate the glassy transition.

Is it possible to verify which of the relations (\ref{e2}) and (\ref{e3}) are
true in our model?
First, let us notice that (\ref{e2}) and (\ref{e3}) are two extremal cases
corresponding to the largest and the smallest excitation energy per surface
area, respectively.
It is thus possible that neither of
them is true and in the glassy phase an intermediate relation holds.
To consider a more general situation let us assume that energy of
excitations scale as $l^{\Theta}$.
In a lattice of the linear size $L$ the number of excitations of size
$l$ scales as $(L/l)^d$ and the total excess energy scales as
$(L/l)^dl^{\Theta}$.
Thus, the excess energy per spin $\delta E$ scales as
\begin{equation}
\delta E\sim l^{\Theta-d}.
\label{e3a}
\end{equation}

To determine $\Theta$ we need a second,
independent measurement of the characteristic length $l$. 
It is already known that $l$ can be also obtained from the
fluctuations of the order parameter using the relation~\cite{SADIQ} 
\begin{equation} 
\chi=1/L^d<1/L^d (\sum_{i} S_i)^2>=l^d,
\label{e4}
\end{equation}
where the magnetization is taken as a corresponding order parameter.
Assuming that relation (\ref{e4}) determines the same 
characteristic length as~(\ref{e3a})
(up to the order of magnitude), we use the following
procedure:  We continuously cool the high-temperature sample down to zero
temperature. 
In this process temperature changes linearly with simulation time
\begin{equation}
T(t)=T_0-rt,
\label{cool}
\end{equation}
where $r$ is the cooling rate.
Then, for the zero-temperature configuration we calculate $\delta E$ and
$\chi$.
Previous Monte Carlo simulations suggest that for model~(\ref{e1}) and $k=2$
the zero-temperature energy excess $\delta E$ decreases with $r$ as  
\begin{equation}
\delta E\sim r^{x_1}, 
\label{e4a}
\end{equation}
where $x_1=0.50(5)$~\cite{LIPDES2}.
Similarly, we expect that $\chi$ also increases as a power
of $r$ 
\begin{equation}
\chi \sim r^{-x_2}. 
\label{e5}
\end{equation}
Inverting (\ref{e5}) and using (\ref{e4}) we obtain $r\sim l^{-d/x_2}$ and from
(\ref{e4a}) we have $\delta E \sim l^{-dx_1/x_2}$.
Finally, comparing the last relation with (\ref{e3a}) we obtain
\begin{equation}
\Theta=d(1-x_1/x_2).
\label{e6}
\end{equation}

To estimate $\Theta$ we made Monte Carlo simulations of model~(\ref{e1})
for $k=2$ and using a sequential Metropolis algorithm~\cite{BINDER}.
Simulations were made for the system size up to $L=70$ and we have checked that
this is sufficient to obtain basically size independent results.
For each cooling rate $r$ we made around 100 independent runs which were used
to calculate $\delta E$ and $1/L^d<1/L^d \sum S_i)^2>$.
The starting temperature was $T=2.8$, which for $k=2$ is above the critical
temperature,
which in this case is $T_c\sim 2.35$~\cite{LIPDESESPRIU}.

\begin{figure} \begin{center}
\epsfig{file=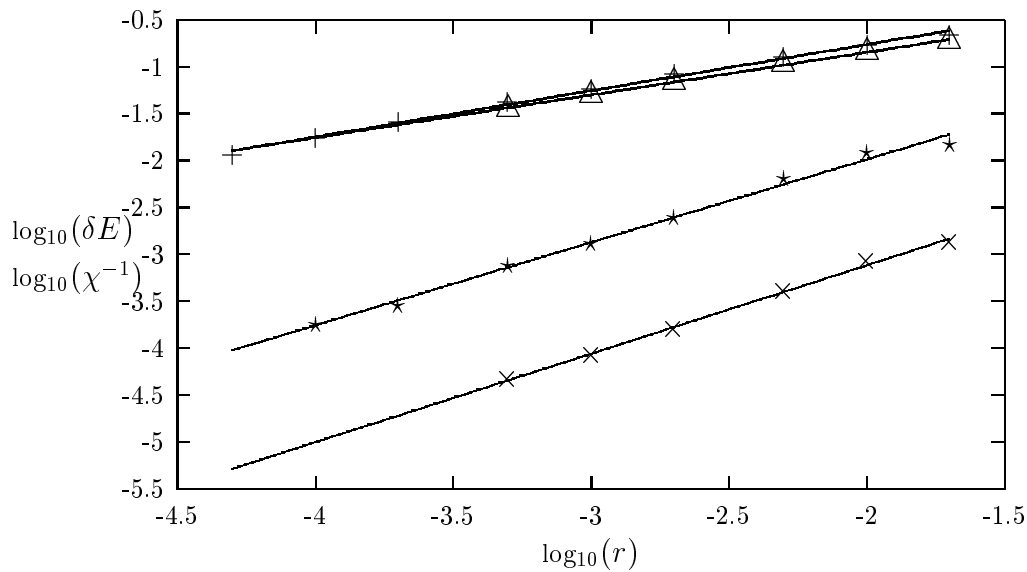}
\end{center}
\caption{
The excess energy $\delta E$ ( +($d=3$) and $\triangle (d=2)$), and $\chi^{-1}$
($\star (d=3)$ and $\times (d=2)$) as a
function of the cooling rate~$r$.
}
\label{f1} 
\end{figure}

The results of our simulations are shown in Fig.~\ref{f1}.
The
relatively good linearity of our data confirm power-law behaviour~(\ref{e4a})
and (\ref{e5}). 
From this data we  estimate $x_1=0.50(5), x_2=0.90(5)$ and using (\ref{e6})
we obtain $\Theta=1.33(5)$.
Such an estimate of $\Theta$ shows that neither (\ref{e2}), which 
corresponds to
$\Theta=2$ nor (\ref{e3}), which corresponds to $\Theta=1$ are correct.
Instead, we have
an intermediate possibility with a noninteger value of
$\Theta$.
Let us notice that since $\Theta<2$ the surface tension of domain walls
vanishes.

For comparison, in Fig.~\ref{f1} we also present results of our simulations
for the two-dimensional (square lattice) version of model~(\ref{e1}) with
$k=2$ (in this case $T_c=0$).
Simulations were made for system sizes up to $L=1000$.
From this data we estimate $x_1=0.46(5)$, $x_2=0.95(5)$ and thus
$\Theta=1.05(5)$. 
It is likely that in this case $\Theta=1$, which would
indicate a trivial nature of domain walls with energy proportional to their
perimeter, the  typical 2D Ising model behaviour.

Although for $k=2$ model~(\ref{e1}) has  slow low-temperature dynamics, it
does
not display a genuine glassy transition.
As we already noted, to model glassy transitions one should 
really study the case of $k=0$. 
However, in this case the  above method encounters some difficulties
since domains of random quench are not only of ferromagnetic type as in
the case of $k=2$, but also antiferromagnetic and even of some mixed types
(see~\cite{LIPDESESPRIU} for some discussion).
For $k=0$ equation~(\ref{e4}) cannot be used and thus, the
exponent $\Theta$  cannot be determined using the above method.

\begin{figure}
\begin{center}
\epsfig{file=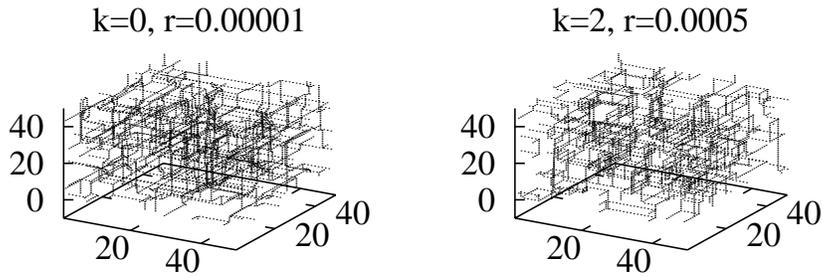,width=15cm}
\end{center}
\caption{
A distribution of unsatisfied plaquettes in the zero-temperature glassy phase.
Simulations were made for the system size $L=100$.
(Only a portion of the system is shown.)
Let us notice that although the cooling for the $k=2$ case is faster it
seems to create larger domains.
It was already suggested that for the case $k=0$ the model should order
much more slowly than for $k=2$~\protect\cite{LIPDESESPRIU}.
}
\label{f2} 
\end{figure}

To get some insight into the $k=0$ case we instead
looked at the distribution of
unsatisfied~\cite{COMM1} plaquettes in the glassy phase (i.e., plaquettes
contributing energy above the ground state).
The random high-temperature sample was slowly cooled down to zero temperature.
Then for the final configuration we located unsatisfied plaquettes
and
their spatial distribution is shown in~Fig.~\ref{f2}.
For comparison we also present similar calculations for the $k=2$ case.
One can see that in both cases energy is concentrated in linear segments.
For $k=2$ this is in agreement with our estimation $\Theta<2$ as
for $\Theta=2$
energy would be localized on two-dimensional surfaces.
Although for $k=0$ we cannot estimate $\Theta$, 
the linear stuctures in Fig.~\ref{f2} strongly suggests
that in this case also $\Theta<2$ and the glassy phase is composed of
tensionless domain walls.

\begin{figure}
\begin{center}
\epsfig{file=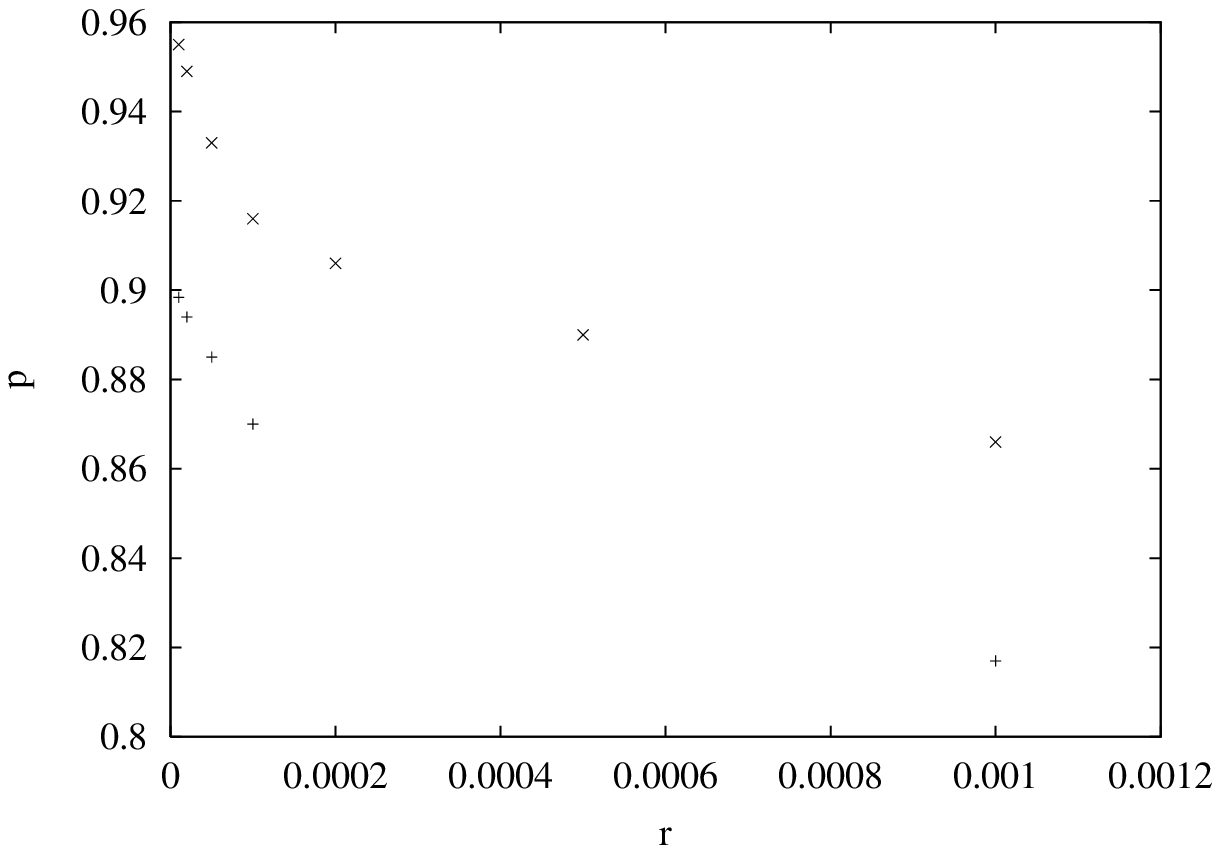,width=10cm,angle=-0}
\end{center}
\caption{
The fraction $p$ of unsatisfied plaquettes which are located in linear
segments as a function of the cooling rate $r$ for $k=0$ (+) and $k=2$
($\times$). } 
\label{f3} 
\end{figure}

Of course, the glassy phase obtained by the finite-rate cooling
contains some regions
other than linear segments where energy is concentrated.
However, we expect that such spots will diminish for decreasing cooling rate
$r$.
To confirm our expectations we measured the ratio $p$ of unsatisfied
plaquettes which belong to linear segments 
compared to the total number of unsatisfied
plaquettes~\cite{COMM2}. 
The results, presented in Fig.~\ref{f3}, show that $p$ indeed increases for
decreasing $r$.
It is also possible that in the limit $r\rightarrow 0$ the fraction $p
\rightarrow 1$. 
Let us notice that a glassy state obtained in such a limit constitutes an ideal
glass~\cite{JACKLE} and the present results might shed some light on this, so 
far hypothetical, state of matter. 
In particular they suggest that in the ideal glass the slow cooling
removes energy-rich spots and leaves only low-energy excitations.

In conclusion, we studied the glassy phase of gonihedric model.
Our result show that energy of excitations in this phase scales as
$l^{\Theta}$ with $\Theta<d-1$.
It confirms earlier expectations that domain walls in this model are
tensionless.
Since the $k=0$ case seems to have a number of properties typical to
ordinary glasses, it would be desirable to check whether this result has
also some analogy in real systems.
\acknowledgements { This work was
partially supported by the KBN grant 5 P03B 032 20, the EC IHP network 
``Discrete Random Geometries: From Solid State Physics to Quantum Gravity''
{\it HPRN-CT-1999-000161} and the ESFnetwork ``Geometry and Disorder: From
Membranes to Quantum Gravity''.} 


\begin{references}    
\bibitem{DROPLET} D.~S.~Fisher and D.~A.~Huse,
Phys.~Rev.~B {\bf 38}, 386 (1988).  
\bibitem{RSB} G.~Parisi,
Phys.~Rev.~Lett.~{\bf 43}, 1754 (1979); Phys.~Rev.~Lett.~{\bf 50}, 1946
(1983).  
\bibitem{MARTIN} F.~Krzakala and O.~C.~Martin, Phys.~Rev.~Lett.~{\bf
85}, 3013 (2000). M.~Pallasini and A.~P.~Young, Phys.~Rev.~Lett.~{\bf 85},
3017 (2000).  
\bibitem {GOTZE} W.~G\"{o}tze, in {\it Liquid, Freezing and Glass
Transition}, Les Houches Summer School, ed. J.~P.~Hansen, D.~Levesque and J.~
Zinn-Justin (North-Holland, Amsterdam, 1989). C.~A.~Angell, Science {\bf 267},
1924 (1995). F.~H.~Stillinger, Science~{\bf 267}, 1935 (1995).
\bibitem {SHORE} J.~D.~Shore, M.~Holzer and J.~P.~Sethna,
Phys.~Rev.~B  {\bf 46}, 11376 (1992).  
\bibitem {LIPDES} A.~Lipowski,
J.~Phys.~A {\bf 30}, 7365 (1997). A.~Lipowski and D.~Johnston, J.~Phys.~A {\bf
33},4451 (2000).   
\bibitem{NEWMAN} M.~E.~J.~Newman and C.~Moore,
Phys.~Rev.~E {\bf 60}, 5068 (1999).  
\bibitem{FRANZ} S.~Franz, M.~Mezard, F.~Ricci-Tersenghi, M.~Weigt, and 
R.~Zecchina, e-print: cond-mat/0103026 
\bibitem{BERNASCONI} J.~Bernasconi, J.~Phys.~(France) {\bf 48}, 559 (1987).
J.~P.~Bouchaud  and M.~M\'{e}zard, J.~Phys.~I (France) {\bf 4}, 1109 (1994).
E.~Marinari, G.~Parisi and F.~Ritort J.~Phys.~A {\bf 27}, 7647 (1994).
\bibitem{LIPDESESPRIU} A.~Lipowski, D.~Johnston and
D.~Espriu, Phys.~Rev.~E {\bf 62}, 3404 (2000).  
\bibitem{SAVVIDY} G.~K.~Savvidy and F.~J.~Wegner, Nucl.~Phys.~B {\bf 413}, 605
(1994). D.~Espriu, M.~Baig, D.~A.~Johnston and R.~P.~K.~C.~Malmini, J.~Phys.~A
{\bf 30}, 405 (1997). R.~V.~Ambartzumian, G.~S.~Sukiasian, G.~K.~Savvidy and
K.~G.~Savvidy, Phys.Lett.~B {\bf  275}, 99 (1992). 
\bibitem {BRAY} A.~J.~Bray, Adv.~in Phys.~{\bf 43}, 357 (1994). 
\bibitem{LIPDES2} A.~Lipowski and D.~Johnston, Phys.~Rev.~E
{\bf 61}, 6375 (2000).   
\bibitem{SWIFT} M.~R.~Swift,
H.~Bokil, R.~D.~M.~Travasso and A.~J.~Bray, Phys.~Rev.~B {\bf 62}, 11494
(2000).  
\bibitem{SADIQ} A.~Sadiq
and K.~Binder, J.~Stat.~Phys.~{\bf 35}, 517 (1984). 
\bibitem {BINDER} See e.g., K.~Binder, in {\it Applications of the Monte Carlo
Method in Statistical Physics}, ed.~K.~Binder, (Berlin: Springer, 1984). 
\bibitem{COMM1} For $k=0$ a plaquette is called unsatisfied if the product of
spin variables on this plaquettes equals 1. Let us notice that for $k=0$ the
coupling in model~(\ref{e1}) is positive while that used e.g.,
in~\cite{LIPDES,LIPDES2} is negative. However, using a simple transformation,
both cases can be made equivalent. 
\bibitem{COMM2} A plaquette separates two elementary cubes.
This plaquette belongs to the linear segment when each of the adjacent
cubes contains only two unsatisfied plaquettes (including the plaquette
under consideration).
\bibitem{JACKLE} J.~J\"ackle,
Rep.~Prog.~Phys.~{\bf 49}, 171 (1986). The idea that in the four-spin model
the 'almost ideal' glassy state might exist was discussed
in~\cite{LIPDES2,SWIFT}.  
\end{references}
\end {document}